\documentclass[twocolumn,showpacs,preprintnumbers,amsmath,amssymb]{revtex4}

\usepackage{graphicx}
\usepackage{dcolumn}
\usepackage{bm}

\begin{document}

\title{Concatenating dynamical decoupling with decoherence-free subspaces for quantum
computation}

\author{Yong Zhang}\email{zhyong98@mail.ustc.edu.cn}\author{Zheng-Wei Zhou}\email{zwzhou@ustc.edu.cn} \author{Bo Yu}\author{Guang-Can Guo}
\address{Key Laboratory of Quantum Information,
University of Science and Technology of China, Chinese Academy of
Sciences, Hefei, Anhui 230026, China}

\begin{abstract}
A scheme to implement a quantum computer subjected to decoherence
and governed by an untunable qubit-qubit interaction is presented.
By concatenating dynamical decoupling through bang-bang (BB) pulse
with decoherence-free subspaces (DFSs) encoding, we protect the
quantum computer from environment-induced decoherence that results
in quantum information dissipating into the environment. For the
inherent qubit-qubit interaction that is untunable in the quantum
system, BB control plus DFSs encoding will eliminate its undesired
effect which spoils quantum information in qubits. We show how
this quantum system can be used to implement universal quantum
computation.
\end{abstract}

\pacs{03.67.Pp, 03.67.Lx}
\maketitle

\section{ Introduction}

Quantum computation (QC) has become a very active field ever since
the discovery that quantum computers can be much more powerful
than their classical counterparts \cite{Shor1,Lloyd,Grover}.
Quantum computers act as sophisticated quantum information
processors, in which calculations are made by the controlled time
evolution of a set of coupled two-level quantum systems. Coherence
in the evolution is essential for taking advantage of quantum
parallelism, which plays an essential role in all quantum
algorithms. However, real physical systems will inevitably
interact with their surrounding environment. No matter how weak
the coupling that prevents an open system from being isolated, the
evolution of the system is eventually plagued by nonunitary
features such as decoherence and dissipation \cite{Gardiner}.
Quantum decoherence, in particular, is a purely quantum-mechanical
effect whereby the system loses its ability to exhibit coherent
behavior by getting entangled with the ambient degrees of freedom.
Decoherence stands as a serious obstacle common to all
applications, including QC, which rely on the capability of
maintaining and exploiting quantum coherence.

Recently, considerable effort has been devoted to designing
strategies able to counteract decoherence. Roughly speaking, three
classes of procedures are available to overcome the decoherence
problem. Two kinds of encoding methods of these strategies in the
field of quantum information are quantum error-correction codes
(QECCs) \cite{Shor2} and decoherence-free subspaces (DFSs, also
called error-avoiding codes) \cite
{Duan1,Duan2,Zanardi1,Zanardi2,Lidar1,Bacon,Kempe,Lidar2}, both
based on encoding the state into carefully selected subspaces of
the Hilbert space of the system. The main difference between the
two encoding strategies is that QECCs is an active strategy, in
which the encoding is performed in such a way that the various
errors are mapped onto orthogonal subspaces so that they can be
diagnosed and reversed, and DFSs instead provide a passive
strategy relying on the occurrence of specific symmetries in the
interaction with the environment, which guarantees the existence
of state space regions inaccessible to noise. The third strategy
can be termed dynamical decoupling or quantum ``bang-bang''(BB)
control \cite{Lorenza1,Lorenza2,Duan3,Vitali1,Zanardi3} after its
classical analog by using strong, fast pulses on quantum systems.
The basic idea is that open-system properties, specifically
decoherence, may be modified if a time-varying control field acts
on the dynamics of the system over time scales that are comparable
to the memory time of the environment. Dynamical decoupling has an
advantage over QECCs and DFSs, because it uses external pulses (BB
pulse) rather than requiring several physical qubits to encode one
logical qubit.

Despite their promise to counteract decoherence in the process of
QC, QECCs and DFSs, in which ancillary physical qubits are
required for protecting quantum information, have their
disadvantage for the construction of a large scale quantum
computer, because the available physical resource is very exiguous
in the present quantum engineering. Dynamical decoupling does not
require an ancillary physical qubit to protect quantum
information, but entirely decoupling system from the environment
requires more complicated pulse operations. Moreover, the inherent
qubit-qubit interaction, which is vital to the implementation of
two-qubit gate, is assumed to be tunable in all the approaches
given above, but this will augment further the complexity of
quantum computer in microstructure. Our effort is devoted to
solving those problems mentioned above. In this work we present an
architecture of quantum computer with fixed coupling between
qubits. In our scheme, by concatenating dynamical decoupling and
DFSs encoding we can simultaneously overcome the effects from
decoherence and qubit-qubit interaction and realize the scalable
fault-tolerant QC.

The structure of the paper is as follows. In Sec. II, we review
dynamical decoupling by BB operations, and we show how to
counteract decoherence via encoding into DFSs and decoupling by BB
operations. In Sec. III we deal with the inherent qubit-qubit
interaction between physical qubits by BB operations. We show in
Sec. IV how the universal QC can be accomplished. Section V is for
discussion and concluding remarks.

\section{Decoherence and Bang-Bang operation}

We consider a two-level quantum system $S$ coupled to an arbitrary
bath $B$, which together form a closed system defined on the
Hilbert spaces
${\cal H}$ $={\cal H}_S
\mathop{\textstyle \otimes }%
{\cal H}_B$ , ${\cal H}_S$ and ${\cal H}_B$ denoting $S$ and $B$
Hilbert spaces, respectively. The dynamics of the quantum system
$S$ coupled to a bath $B$ evolves unitarily under the Hamiltonians
\begin{equation}
H=H_S%
\mathop{\textstyle \otimes }%
I_B+I_S%
\mathop{\textstyle \otimes }%
H_B+H_{SB},
\end{equation}
where $H_S$, $H_B$, and $H_{SB}$ are the system, bath, and
interaction Hamiltonians, respectively. The interaction
Hamiltonians between the system and bath can be written as
\begin{equation}
H_{SB}=\sigma _x%
\mathop{\textstyle \otimes }%
b_x+\sigma _y%
\mathop{\textstyle \otimes }%
b_y+\sigma _z%
\mathop{\textstyle \otimes }%
b_z.  \label{e1}
\end{equation}
Here the $\sigma_\alpha$'s ($\alpha=x,y,z$) are the
spin-$\frac{1}{2}$ Pauli operators on physical qubit and the
$b_\alpha$'s are operators on the degrees of freedom of
environment. Due to the interaction Hamiltonian, the quantum
system will entangle with the environment so that the quantum
information encoded into quantum states irreversibly dissipates
into the environment, this is the so-called decoherence. The
objective of dynamical decoupling with BB operations used in our
scheme is to modify this unwanted evolution.

The process of dynamical decoupling by BB operations, which
counteracts decoherence by applying sequences of strong and fast
pulses, serves for protecting the evolution of $S$ against the
effect of the interaction $H_{SB}.$ In the standard view of the
dynamical decoupling, a set of realizable BB operations can be
chosen such that they form a discrete (finite order) subgroup of
the full unitary group of operation on the Hilbert space of the
system. Denote the subgroup ${\cal G}$ and its elements $g_k$,
$k=0,1,\ldots ,\left| {\cal G}\right| -1$, where $\left| {\cal
G}\right| $ is the order of the group. The cycle time is
$T_c=\left| {\cal G}\right| \Delta t$, where $\left| {\cal
G}\right| $ is now also the number of pulse operations, and
$\Delta t$ is the time that the system evolves freely between
operations under $U_0=\exp (-iHt)$. The evolution of the system
now is given by
\begin{equation}
U(T_c)=%
\mathop{\textstyle \prod }%
_{k=0}^{\left| {\cal G}\right| -1}g_k^{\dagger}U_0(\Delta
t)g_k\equiv e^{iH_{eff}T_c}
\end{equation}
$H_{eff}$ denotes the resulting effective Hamiltonian. Obviously,
to satisfy the above equation, it is required that the pulses in
the sequence are very fast and strong compared with the evolution
of Hamiltonian $H$, which is the origin of the name ``bang-bang''
operation. In this BB limit, the system will evolve under the
effective Hamiltonian
\begin{equation}
H\rightarrow H_{eff}=\frac 1{\left| {\cal G}\right| }%
\mathop{\textstyle \sum }%
_{k=0}^{\left| {\cal G}\right| -1}g_k^{\dagger}Hg_k\equiv
\mathop{\textstyle \prod }%
\nolimits_{{\cal G}}(H).
\end{equation}
The map $%
\mathop{\textstyle \prod }%
_{{\cal G}}$ commutes with all $g_k$ so that the action of the
controller over times longer than the averaging period $T_c$ only
preserves the set of operators which are invariant under ${\cal
G}$, thereby enforcing a ${\cal G} $ symmetrization of the
evolution of $S$ \cite{Viola}. Recently, a general result has been
established by Facchi \emph{et al}. \cite{Facchi}, which states
that dynamical decoupling can be accomplished by a sequence of
arbitrary (fast and strong) pulses and symmetry or group structure
is not necessary, and the above procedure of decoupling by
``symmetrization'' arises as a special case. The main drawback of
BB pulse decoupling procedures is that the timing constraints are
particularly stringent. In fact, perfect decoupling from the
environment is obtained only in the infinitely fast control limit
\cite {Lorenza2,Vitali1,Vitali2}, but it has been established that
these decoupling schemes can be effective in a realistic situation
with control pulses with finite strength and time duration
\cite{Vitali1,Vitali3}.

Now let us first present our approach to counteract decoherence.
For modifying the coupling induced by the Hamiltonian in Eq.
(\ref{e1}), we consider a single BB operation
$U_{z1}=\exp(-i\sigma_z\pi/2)=-i\sigma_z$, and when no pulses are
applied the unit operator $I$ denotes the operation on qubits.
Using the commutation relation for Pauli operators, we have

\begin{equation}
U^\dagger_{z1}\sigma _xU_{z1}=\sigma _z\sigma _x\sigma _z=-\sigma
_x,
\end{equation}
\begin{equation}
U^\dagger_{z1}\sigma _yU_{z1}=\sigma _z\sigma _y\sigma _z=-\sigma
_y,
\end{equation}
\begin{equation}
U^\dagger_{z1}\sigma _zU_{z1}=\sigma _z\sigma _z\sigma _z=\sigma
_z.
\end{equation}
Thus after cycles of BB operations, we can obtain the effective
interaction Hamiltonian
\begin{equation}
H_{SB}\rightarrow
\mathop{\textstyle \prod }%
\nolimits(H_{SB})=\sigma _z%
\mathop{\textstyle \otimes }%
b_z,
\end{equation}
which still introduces phase decoherence. In order to counteract
phase decoherence, we can encode quantum information into DFSs. We
use a well-known code \cite{Duan1,Duan2,Palma} which two physical
qubits encode a logical qubit,
\begin{equation}
\left| 0\right\rangle_L =\left| 0_11_2\right\rangle%
 and \left| 1\right\rangle_L =\left|
1_10_2\right\rangle. \label{e2}
\end{equation}
Here $i=1,2$ indexes physical qubits. For the system consisting of
two physical qubits, the BB operation on the two physical qubits,
correspondingly, can be defined as collective rotation:
$U_z=U_{z1}\otimes
U_{z2}=\exp(-i\sigma_1^z\pi/2)\otimes\exp(-i\sigma_2^z\pi/2)=-\sigma
_1^z\otimes \sigma _2^z$,and then $
\mathop{\textstyle \prod }%
\nolimits(H_{SB})=(\sigma_1^z+\sigma_2^z)%
\mathop{\textstyle \otimes }%
b_z$. Clearly, such encoding on a pair of physical qubits ensures
that the encoded states are decoherence-free for phase error only
if the disturbances from the environment around the system are
identical. In other words, the two qubits must be arranged so
close to each other that they undergo collective phase
decoherence. Here the DFSs encoding together with BB operations
serve for combating decoherence.

In Refs. \cite{Byrd,lidar3}, Byrd and Lidar have proposed a
comprehensive encoding and decoupling solution to problems of
decoherence. Decoherence is first reduced by encoding a logical
qubit into two qubits, then completely eliminated by an efficient
set of decoupling pulse sequences, in which cycles of pairs of BB
pulses generated from the same exchange Hamiltonian are used to
eliminate errors other than dephasing. The quantum code in our
scheme is analogous to the one they have proposed for reducing
phase decoherence. Then we apply directly a kind of simple BB
pulse on a physical qubit to selectively decouple the system from
the environment, which reduces the complexity of pulse operation.
In our scheme untunable qubit-qubit interaction can be controlled
by BB operations as discussed in the following section.

\section{Interaction and Bang-Bang operation}

To realize QC, any universal quantum gates (quantum operations)
must include single-qubit gates and two-qubit gates. A traditional
way for the implementation of single-qubit and two-qubit gates
requires a control on two qubits level that is an ability to
``switch on'' and to ``switch off'' interaction between qubits.
But an ``always on'' coupling can cause certain problems for
quantum information preservation and QC. For example, if the
interaction between two physical qubits in the code (\ref{e2}) is
Heisenberg exchange interaction \cite{Ruskai}, the computational
basis will always be flipped under the exchange Hamiltonian, which
spoils quantum information in qubits. In general, quantum
computers exploit control techniques \cite {Kane,Vrijen} to tune
the interaction between two physical qubits to avoid the undesired
effect of the coupling, and tunability of the interaction constant
is at the heart of many solid-state proposals, but this prove
extremely difficult to achieve experimentally. Recently, some
schemes of QC governed by always on interaction have been
presented \cite{Zhou,Benjamin,Zhang}. In our scheme, we discuss
the case that the interaction is always on and untunable, and we
exploit BB operations to selectively decouple two physical qubits.

Now we consider the general exchange interaction between physical
qubits. The exchange interaction Hamiltonian in the system has the
form
\begin{equation}
H_I=J_x\sigma _1^x%
\mathop{\textstyle \otimes }%
\sigma _2^x+J_y\sigma _1^y%
\mathop{\textstyle \otimes }%
\sigma _2^y+J_z\sigma _1^z%
\mathop{\textstyle \otimes }%
\sigma _2^z, \label{e3}
\end{equation}
where $J_a$'s, $(a=x,y,z)$ are exchange interaction constants.

We first consider the case of a single logical qubit. Under the
self-exchange interaction, we find
\begin{equation}
H_I\left| 0\right\rangle _L=(J_x+J_y)\left| 1\right\rangle
_L-J_z\left| 0\right\rangle _L,
\end{equation}
\begin{equation}
H_I\left| 1\right\rangle _L=(J_x+J_y)\left| 0\right\rangle
_L-J_z\left| 1\right\rangle _L.
\end{equation}
Obviously, quantum information encoded will be spoiled by the
self-exchange interaction. We selectively decouple the two
physical qubits encoded into a logical qubit by introducing a
selective decoupling BB operation
$R_z=I_1\otimes\exp(-i\sigma_2^z\pi/2)=-iI_1\otimes\sigma_2^z$. We
obtain
\begin{equation}
R_z^{\dagger}\sigma _1^x%
\mathop{\textstyle \otimes }%
\sigma _2^xR_z=\sigma _1^x%
\mathop{\textstyle \otimes }%
\sigma _2^z\sigma _2^x\sigma _2^z=-\sigma _1^x%
\mathop{\textstyle \otimes }%
\sigma _2^x,
\end{equation}
\begin{equation}
R_z^{\dagger}\sigma _1^y%
\mathop{\textstyle \otimes }%
\sigma _2^yR_z=\sigma _1^y%
\mathop{\textstyle \otimes }%
\sigma _2^z\sigma _2^y\sigma _2^z=-\sigma _1^y%
\mathop{\textstyle \otimes }%
\sigma _2^y,
\end{equation}
\begin{equation}
R_z^{\dagger}\sigma _1^z%
\mathop{\textstyle \otimes }%
\sigma _2^zR_z=\sigma _1^z%
\mathop{\textstyle \otimes }%
\sigma _2^z\sigma _2^z\sigma _2^z=\sigma _1^z%
\mathop{\textstyle \otimes }%
\sigma _2^z.
\end{equation}
So after cycles of BB operations, we obtain effective self-interaction $%
\mathop{\textstyle \prod }%
(H_I)=J_z\sigma _1^z%
\mathop{\textstyle \otimes }%
\sigma _2^z$, which is equivalent to Ising interaction; the encoded states $%
\left| 0_L\right\rangle $ and $\left| 1_L\right\rangle $ in Eq.
(\ref{e2}) are degenerate under the effective self-interaction.
Therefore, if we store information in these states, no evolution
whatsoever is present. In other words, for the untunable exchange
interaction quantum information is stabilized by means of BB
control and quantum encoding.

Until now, we have introduced two BB operationS $U_z$ and $R_z$.
As already noted, the two BB operations are used on qubitS 1 and 2
to counteract decoherence and undesired interaction. Actually, the
pulse operations $R_z=I_1\otimes\exp(-i\sigma_2^z\pi/2)$ only act
on physical qubit 2. For physical qubit 1, only the pulse
operation $\sigma_1^z$ has an effect on the decoherence. But there
are two kinds of pulse operations in $U_z$ and $R_z$ effected on
qubit 2 to selectively eliminate not only qubit-qubit interaction
but also qubit-environment interaction. In other words, the number
of pulse operations on qubits 1 and 2 is dissimilar. Because we
apply the same pulse operations ($\sigma^z$) on every physical
qubit, the time intervals $\Delta t_1$ on qubit 1 and $\Delta t_2$
on qubit 2 are different too. This implies that we have applied a
kind of nonsynchronous pulse operations to overcome
environment-induced decoherence and unwanted coupling between
physical qubits.

Let us now show how to devise nonsynchronous pulse operations for
decoupling different interactions. We can elaborately devise a set
of programmed pulse operations in which the time intervals of the
BB operations on two qubits are varied according to the program.
In our scheme, unitary pulse operations are $U_z$ and $R_z$ as
given above. Here we assume that the BB operation $U_z$ begins at
time ${t}_0 $=0 and devise the time interval between two pulse
operations is constant $\Delta t$. Then we devise the BB operation
$R_z$ begins at time ${t}_0+\Delta t/2$ and the time interval is
$\Delta t$ too. So the time intervals between a pair of pulses on
qubits 1 and 2 have the relation $\Delta t_1=2\Delta t_2$. In fig.
1 we focus on the evolution of the $y$ ingredient in Hamiltonian
$H_{SB}$ under the cycles of BB pulses. (The same conclusion
adapts to the $x$ ingredient in $H_{SB}$.) $T_1=2\Delta t_1$ and
$T_2=2\Delta t_2$ denote the cycle time of decoupling operations
on qubits 1 and 2, respectively. After cycles of pulse operations,
the total effect of error operators ($Y$ in the figure) on qubits
1 and 2, respectively, is zero in the cycles time $NT_i$
($i=1,2$), here $N$ and $N_i (i=1,2,3)$ given in the following are
positive integer. This implies that decoherence on qubits 1 and 2
is hold back. In addition, by similar analysis, we find that for
the self-interaction between qubits 1 and 2, the total effect of
the error operator
$J_x\sigma^x_1\otimes\sigma^x_2+J_y\sigma^y_1\otimes\sigma^y_2$ is
also eliminated in the cycles time $T=N_1T_1=N_2T_2$, so in $y$
axis qubits 1 and 2 are decoupled. The result shows that the
programmed BB pulse operations can eliminate or selectively
eliminate not only qubit-environment interaction but also
qubit-qubit interaction. This gives us a very heuristic solution
to elimination of undesired coupling. The method of decoupling
with programmed unsymmetrical pulse operations may be of great
benefit to the implementation of QC in many complicated
circumstances.

In the above discussion, we present a dynamical decoupling scheme
based on group averaging formulation. It is noteworthy that for
the two-qubit system the operation set $\{I,U_z,R_z\}$ has no
group structure, which accords with the result of Ref.
\cite{Facchi}.

 We still need to show how the interaction between two logical
qubits influences the encoded states of logical qubits. The
exchange interaction in Eq. (\ref{e3}) between two logical qubits
will induce unwanted  flow of quantum information between two
logical qubits. This will inevitably result in the failure of the
preservation of quantum information and QC. In our scheme, quantum
computer is constructed in  a one-dimensional array of physical
qubits. Now, we introduce new logical qubits $L_2$ and $L_3$ (See
Fig. 2). For logical qubit $L_2$, two selective decoupling BB
operations are chosen as $U_x$ and $R_x$, here $U_x=U_{x3}\otimes
U_{x4}=\exp(-i\sigma_3^x\pi/2)\otimes\exp(-i\sigma_4^x\pi/2)=-\sigma
_3^x\otimes \sigma _4^x$ and $R_x=I_3\otimes
U_{x4}=-iI_3\otimes\sigma_4^x$. Then, we can obtain the effective
interaction Hamiltonian
$
\mathop{\textstyle \prod }%
\nolimits(H_{SB})=(\sigma_3^x+\sigma_4^x)%
\mathop{\textstyle \otimes }%
b_x$
and the effective self-interaction $%
\mathop{\textstyle \prod }%
(H_I)=J_x\sigma _3^x%
\mathop{\textstyle \otimes }%
\sigma _4^x$. Accordingly, two encoded states of $L_2$ encoded in
DFS can be written as
\begin{equation}
\left| 0\right\rangle_{L_B} =\frac12(\left|
0_3\right\rangle+\left| 1_3\right\rangle)(\left|
0_4\right\rangle-\left| 1_4\right\rangle)\label{e4},
\end{equation}
\begin{equation}
\left| 1\right\rangle_{L_B} =\frac12\left| 0_3\right\rangle-\left|
1_3\right\rangle)(\left| 0_4\right\rangle+\left| 1_4\right\rangle)
\label{e5},
\end{equation}
where the subscript $B$ denotes the method of decoupling and
encoding for logical qubit $L_2$. Similarly, two selective
decoupling subgroups of logical qubit $L_3$ are chosen as $U_y$
and $R_y$, here $U_y=U_{y5}\otimes
U_{y6}=\exp(-i\sigma_5^y\pi/2)\otimes\exp(-i\sigma_6^y\pi/2)=-\sigma
_5^y\otimes \sigma _6^y$ and $R_y=I_5\otimes
U_{y6}=-iI_5\otimes\sigma_6^y$, and then, the quantum code in DFS
will have the form

\begin{equation}
\left| 0\right\rangle_{L_C} =\frac12(\left|
0_5\right\rangle+i\left| 1_5\right\rangle)(\left|
0_6\right\rangle-i\left| 1_6\right\rangle),
\end{equation}
\begin{equation}
\left| 1\right\rangle_{L_C} =\frac12(\left|
0_5\right\rangle-i\left| 1_5\right\rangle)(\left|
0_6\right\rangle+i\left| 1_6\right\rangle).
\end{equation}
Obviously, with selective decoupling and encoding into DFSs, $L_2$
and $L_3$ can overcome decoherence and unwanted internal
interaction as $L_1$ does .

Now, we focus on the coupling between logical qubits $L_1$ and
$L_2$ that is equivalent to the coupling between physical qubits 2
and 3. The inherent interaction Hamiltonian between qubit 2 and 3
has the form as shown in Eq. (\ref{e3}). For physical qubit 2, the
pulse operation is $\sigma_2^z$, then the evolution of the $x$ and
$y$ ingredients in Hamiltonian $H_{SB}$ is changed. For qubit 3,
the pulse operation is $\sigma_3^x$ which changes the evolution of
the $y$ and $z$ ingredients in Hamiltonian $H_{SB}$. Then, after
cycles of pulse operations in the time $T=N_2 2\Delta t_2=N_3
2\Delta t_3$, we obtain $\prod(\sigma_2^x\otimes\sigma_3^x)=0$ and
$\prod(\sigma_2^z\otimes\sigma_3^z)=0$. So the evolution of the
$x$ and $z$ ingredients in Hamiltonian $H_{SB}$ is eliminated. As
far as the evolution of the $y$ ingredient is concerned, since
pulses effect on qubit 2 at the interval of $\Delta t_2$, but on
qubit 3 at the interval of $\Delta t_1$, here $\Delta t_1=2\Delta
t_2$, the evolution about $y$ axis on qubits 2 and 3 is
unsymmetrical, then $\prod(\sigma_2^y\otimes\sigma_3^y)=0$, i.e.,
the evolution of the $y$ ingredient in Hamiltonian $H_{SB}$ is
eliminated. This can also be illuminated by Fig. 1. To sum up,
with cycles of pulse operations, the effect of Hamiltonian
$H_{SB}$ between qubits 2 and 3 is eliminated. In other words,
$L_1$ is entirely decoupled from $L_2$. The same conclusion can be
drawn for logical qubits $L_2$ and $L_3$.

We showed above that with BB pulse operations and quantum encoding
into DFS, the three logical qubits overcome not only
environment-induced decoherence but also unwanted inherent
interaction which is always on and untunable between physical
qubits. And we devise that the three logical qubits are effected
with three different BB operations so that every logical qubit is
decoupled from others. Then, we can construct a scalable quantum
computer with the three logical qubits as a unit of computation,
i.e., the quantum computer has the periodic structure
$AABBCCAABBCC\cdots$, where $AA$, $BB$, and $CC$ denote encoded
logical qubits analogous to $L_1$, $L_2$,and $L_3$, respectively.

\begin{figure}[here]

\includegraphics[width=8cm]{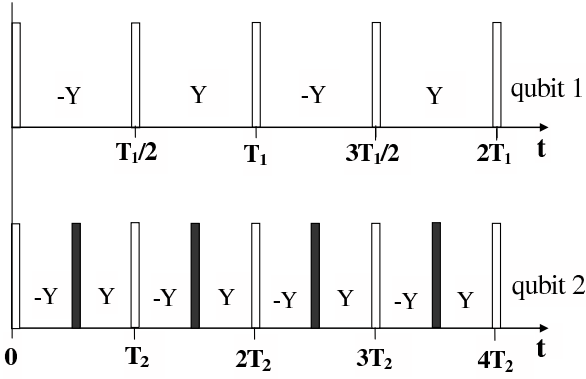}

\caption{the evolution of physical qubit 1 and 2 about the $y$
axis under Hamiltonian $H_{SB}$ and pulse operations. White and
black rectangles denote strong and fast pulse operations $U_z$ and
$R_z$, respectively. $T_1$ and $T_2$ denote the cycle time of
decoupling operations on qubit 1 and 2. $Y$=$\sigma_y$ is error
operator on physical qubit.}
\end{figure}

\begin{figure}[here]
\includegraphics[width=8cm]{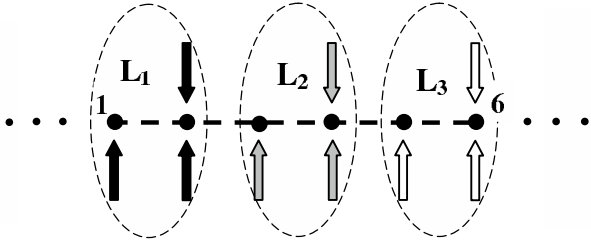}

\caption{Architecture of three logical qubits in the quantum computer.%
 Each dot is a physical qubit and the dashed lines represent
interaction between qubits. Every logical qubit consists of two
physical qubits. Arrows with different colors denote different
pulse operations on logical qubit.}
\end{figure}

\section{Quantum Computation}

Our discussion so far has concentrated on the preservation of
quantum information. To carry out quantum information, we must
have the ability to manipulate encoded quantum information. Thus
we still need to show that universal QC can actually be performed
in our scheme. DiVincenzo shows that for any unitary
transformation on quantum states it is sufficient to apply (a) all
single-qubit rotations [SU(2)] together with (b) the two-qubit
controlled-NOT (CNOT) gate on any two logical qubits
\cite{DiVicenzo}.

In our scheme, we assume that any single-qubit operations on
physical qubits are realizable at will by virtue of external
pulses. We can define logical operations (denoted by a bar) which
act on the encoded qubits. For example, $\overline{X}:\left|
0_L\right\rangle \leftrightarrow \left| 1_L\right\rangle $. For
logical qubit $L_1$, $\overline{X}=(J_x\sigma _1^x\otimes\sigma
_2^x+J_y\sigma _1^y\otimes\sigma _2^y)/(J_x+J_y)$. Logical
$\overline{X}$ operation can be easily achieved by recoupling
qubits 1 and 2 with the interaction Hamiltonian as shown in Eq.
(\ref{e3}). We adjust the time intervals of pulses on qubits 1 and
2 both to $\Delta t_3$, where $\Delta t_3=\Delta t_2/2=\Delta
t_1/4$. In other words, only synchronous collective BB pulses are
applied, which just eliminate the coupling from environment but
have no effect on qubit-qubit coupling $H_I$. Then, we have

\begin{eqnarray}
e^{i\theta H_I}\left| i\right\rangle _L &=&e^{i\theta (J_x\sigma _1^x%
\mathop{\textstyle \otimes }%
\sigma _2^x+J_y\sigma _1^y%
\mathop{\textstyle \otimes }%
\sigma _2^y+J_z\sigma _1^z%
\mathop{\textstyle \otimes }%
\sigma _2^z)}\left| i\right\rangle _L  \nonumber \\
&=&e^{-i\theta J_Z}e^{i\theta (J_x+J_y)\overline{X}}\left|
i\right\rangle _L.
\end{eqnarray}
By the free evolution under the inherent interaction Hamiltonian,
we can easily accomplish logical $\overline{X}$ operation. We must
note that the time intervals of pulses on qubits 2 and 3 are still
unequal; this implies that after cycles of pulse operations in the
time $T=N_2 2\Delta t_3=N_3 2\Delta t_1$, qubit 2 remains
decoupled from qubit 3; logical $\overline{X}$ operation on $L_1$
therefore has no impacts on other logical qubits. We can also
implement logical $Z$ operation $\overline{Z}=(\sigma _1^z-\sigma
_2^z)/2$ by direct pulse operations on physical qubits, then
$\overline{X}$ and $\overline{Z}$ generate all encoded-qubit SU(2)
transformations.

By inspection of quantum codes of logical qubits $L_1$, $L_2$, and
$L_3$, we find that the DFSs of $L_2$ and $L_3$ can be obtained by
performing a unitary transformation on that of $L_1$. For example,
the transformation of DFSs between $L_1$ and $L_2$ is a Hadamard
transformation. Obviously, single-encoded-qubit operations, which
preserve the DFSs of $L_1$ and $L_2$, respectively, have the same
unitary transformation too. Then, by performing a transformation
on single-encoded-qubit gate given above, all single-encoded-qubit
operations [SU(2)] on $L_2$ and $L_3$ can be easily achieved (See
Table I).

Two-encoded-qubit CNOT gate seems to be more complicated, but in
our scheme it is very easy to accomplish the two-qubit gate. For
the convenience of discussion, let us assume that we want to do a
CNOT operation from logical qubit $L_1$ to $L_2$ in Fig. 2. To
obtain a two-qubit gate, we consider the imprimitive gate
$W=e^{i\theta \sigma ^z\otimes \sigma ^z}$, which is equivalent to
a controlled rotation about the $z$ axis \cite{Bremmer},

\begin{equation}
e^{i\theta \sigma ^z\otimes \sigma ^z}\equiv \left| 0\right\rangle
\left\langle 0\right|
\mathop{\textstyle \otimes }%
I+\left| 1\right\rangle \left\langle 1\right|
\mathop{\textstyle \otimes }%
e^{i2\left| \theta \right| \sigma ^z}
\end{equation}
Conjugated by single-qubit Hadamard  $H=%
{\displaystyle {1 \over \sqrt{2}}}%
\left[
\begin{array}{cc}
1 & 1 \\
1 & -1
\end{array}
\right] $ operation on the second qubit, $W$ can be used to
implement a CNOT:

\begin{equation}
\text{CNOT}\equiv \left| 0\right\rangle \left\langle 0\right|
\mathop{\textstyle \otimes }%
I+\left| 1\right\rangle \left\langle 1\right|
\mathop{\textstyle \otimes }%
e^{i(\frac \pi 2)\sigma ^x}
\end{equation}
To implement a encoded CNOT between $L_1$ and $L_2$, we must
recouple the two logical qubits with an interaction in the form
$\overline{Z}_{L_1}\otimes\overline{Z}_{L_2}$. We perform a
unitary Hadamard transformation on $L_2$. In other words, we
change the BB pulses characterized by $-\sigma _3^x\otimes \sigma
_4^x$ and $-iI_3\otimes \sigma _4^x$ to the same with $L_1$, and
the quantum code in Eqs. (\ref{e4}) and (\ref{e5}) to the same
with that in Eq. (\ref{e2}), i.e.,
$|0\rangle_{L_B}\rightarrow|0\rangle_{L_A}=|0_31_4\rangle$ and
$|1\rangle_{L_B}\rightarrow|1\rangle_{L_A}=|1_30_4\rangle$. It
should be noted that $L_2$ and $L_3$ are still entirely decoupled
after the unitary transformation. Then the effective interactions
between qubits 1, 2, 3 and 4 all are in the form of Ising
interaction. In this system we assume that the interaction only
exists between any nearest-neighbor physical qubits. Obviously,
$\overline{Z}_{L_1}\otimes\overline{Z}_{L_2}=\sigma_2^z\otimes
\sigma_3^z$, two-encoded-qubit CNOT gate can be implemented by the
evolution under the effective interaction $\sigma_2^z\otimes
\sigma_3^z$ and single-qubit Hadamard operation conjugately
effected on a physical qubit.  Similarly, we can implement CNOT
operation between $L_2$ and $L_3$.

As above, we showed that it is possible to perform all single- and
two-encoded-qubit operations by means of pulse operations and
evolution under inherent interaction. In our scheme, single- and
two-encoded-qubit operations do not influence decoupling
operations and preserve DFSs all the time, so quantum states
encoded with quantum information will not undergo decoherence,
then we implement universal, fault-tolerant QC.

\section{Discussion and Conclusion}

In this paper we have presented a scheme of scalable quantum
computer governed by untunable exchange Hamiltonian. We combine
ideas from the theory of decoherence-free subspaces and BB control
to solve the problem of strong decoherence. Cycles of simple BB
pulses are used to selectively decouple the system from external
environment, then by encoding two physical qubits into a DFS, we
obtain full protection against strong decoherence. By
concatenating BB control with the DFSs encoding, our scheme
decreases the number of physical qubits required to counteract
decoherence. It is highly important for the physicist to reduce
the physical resource needed for implementation of scalable
quantum computer, because quantum computing resources available
are still a stringent requirement for practical quantum
engineering. Comparing with other decoupling scheme, in our scheme
only very simple BB pulses are applied which is easy to
accomplish.

Furthermore, we have discussed the influence of an always on and
untunable interaction between physical qubits on the logical
qubits. The undesired effects of the internal interaction can be
eliminated via cycles of BB operations, which simplifies the
physical structure of quantum computer that is devised in a very
complicated manner for implementing the tunability of the coupling
strength in many QC proposals. By different unsymmetrical
decoupling operations, every logical qubit is entirely decoupled
from others. With direct pulse operations on physical qubits and
effective interaction, we can achieve all single- and
two-encoded-qubit gates for implementing universal QC. Moreover,
in our scheme all single- and two-encoded-qubit operations
preserve logical qubits in a DFS all the time, so we implement
universal, fault-tolerant QC.

\section{ Acknowledgments}

This work was funded by National Fundamental Research Program
(2001CB309300), National Natural Science Foundation of China under
Grants No.10204020, the Innovation funds from Chinese Academy of
Sciences, and also by the outstanding Ph. D thesis award and the
CAS's talented scientist award entitled to Luming Duan.

\newpage
\begin{table*}

\renewcommand{\arraystretch}{2.0}
\caption{Comparison of properties between logical qubit $L_1$,
$L_2$ and $L_3$.}
\begin{tabular}{ccccc}  \hline\hline& $L_1$ & $L_2$ & $L_3$ \\\hline\hline
$U_\alpha$ & $-\sigma _1^z\otimes \sigma _2^z$ & $-\sigma
_3^x\otimes \sigma _4^x$ & $-\sigma _5^y\otimes \sigma _6^y$ \\
\hline $R_\alpha$ & $-iI_1\otimes \sigma _2^z$ &
$-iI_3\otimes \sigma _4^x$ & $-iI_5\otimes \sigma _6^y$ \\
\hline $\left| 0\right\rangle _L$ & $\left| 0_11_2\right\rangle $
& $\frac 12(\left| 0_3\right\rangle +\left| 1_3\right\rangle
)(\left| 0_4\right\rangle -\left| 1_4\right\rangle )$ & $\frac
12(\left| 0_5\right\rangle +i\left| 1_5\right\rangle )(\left|
0_6\right\rangle -i\left| 1_6\right\rangle )$ \\ \hline $\left|
1\right\rangle _L$ & $\left| 1_10_2\right\rangle $ & $\frac
12(\left| 0_3\right\rangle -\left| 1_3\right\rangle )(\left|
0_4\right\rangle +\left| 1_4\right\rangle )$ & $\frac12(\left|
0_5\right\rangle -i\left| 1_5\right\rangle )(\left|
0_6\right\rangle +i\left| 1_6\right\rangle )$
\\ \hline $\overline{X}$ & $(J_x\sigma _1^x\otimes \sigma
_2^x+J_y\sigma _1^y\otimes \sigma _2^y)/(J_x+J_y)$ & $(J_y\sigma
_3^y\otimes \sigma _4^y+J_z\sigma _3^z\otimes \sigma
_4^z)/(J_y+J_z)$ & $( J_x\sigma _5^x\otimes \sigma _6^x+J_z\sigma
_5^z\otimes \sigma _6^z)/( J_x+J_z)$ \\ \hline $\overline{Z}$ &
$(\sigma _1^z-\sigma _2^z)/2$ & $(\sigma _3^x-\sigma _4^x)/2 $ &
$(\sigma _5^y-\sigma _6^y)/2$\\\hline\hline
\end{tabular}
\end{table*}


\begin{references}

\bibitem{Shor1}  P. W. Shor, in Proceedings of the Thirty-Fifth Annual Symposium
on Foundations of Computer Science. Edited be S.Goldwsser(IEEE
Computer Society, New York, 1994), pp. 124-134

\bibitem{Lloyd}  S. Lloyd, Science {\bf 273}, 1073(1996)

\bibitem{Grover}  L. K. Grover, Phys. Rev. Lett. {\bf 79}, 325 (1997)

\bibitem{Gardiner}  C. W. Gardiner, Quantum Noise (Springer, Berlin, 1991)

\bibitem{Shor2}  P. W. Shor, Phys. Rev. A {\bf 52}, R2493 (1995); A. M. Steane,
Phys. Rev. Lett. {\bf 77}, 793 (1996); E. Knill and R. Laflamme, Phys. Rev. A {\bf %
55}, 900 (1997).

\bibitem{Duan1}  L. M. Duan and G. C. Guo, Phys. Rev. Lett. {\bf 79}, 1953 (1997).

\bibitem{Duan2}  L. M. Duan and G. C. Guo, Phys.R ev. A {\bf 57}, 737 (1998).



\bibitem{Zanardi1}  P. Zanardi and M. Rasetti, Mod. Phys. Lett. B {\bf11},
1085 (1997).

\bibitem{Zanardi2}  P. Zanardi and M. Rasetti,Phys. Rev. Lett.{\bf 79}%
,3306 (1997).

\bibitem{Lidar1}  D. A. Lidar, I. L. Chuang, and K. B. Whaley, Phys. Rev. Lett. {\bf 81}, 2594 (1998).

\bibitem{Bacon}  D. Bacon, J. Kempe, D. A. Lidar, and K. B. Whaley, Phys.
Rev. Lett. {\bf 85}, 1758 (2000).

\bibitem{Kempe}  J. Kempe, D. Bacon, D. A. Lidar, and K. B. Whaley, Phys. Rev. A
{\bf 63}, 042307 (2001).

\bibitem{Lidar2}  D. A. Lidar, D. Bacon, J. Kempe, and K. B. Whaley, Phys. Rev. A
{\bf 63}, 022306 (2001).

\bibitem{Lorenza1}  L. Viola and S. Lloyd, Phys. A {\bf 58}, 2733 (1998).

\bibitem{Lorenza2}  L. Viola, E. Knill, and S. Lloyd, Phys. Rev. Lett. {\bf 82},
2417 (1999).

\bibitem{Duan3}  L. M. Duan and G. C. Guo, Phys. Lett. A {\bf 261}, 139 (1999).

\newpage

\bibitem{Vitali1}  D. Vitali and P. Tombesi, Phys. Rev. A {\bf 59}, 4178 (1999).

\bibitem{Zanardi3}  P. Zanardi, Phys. Lett. A {\bf 258}, 77 (1999).

\bibitem{Viola}  L. Viola, E. Knill, and S. Lloyd, Phys. Rev. Lett.{\bf 85}, 3520 (2000).

\bibitem{Facchi} P. Facchi, D. A. Lidar, and S. Pascazio, e-print
quant-ph/03030132

\bibitem{Vitali2}  D. Vitali and P. Tombesi, Phys. Rev. A{\bf 65}, 012305 (2002).

\bibitem{Vitali3}  D. Vitali, J. Opt. B: Quant Semi. Opt. {\bf 4}, 337 (2002).

\bibitem{Palma}  G. M. Palma, K. A. Suominen, and A. K. Ekert, Proc. R. Soc. London,
Ser. A {\bf 452}, 567 (1996).

\bibitem{Byrd}  M. S. Byrd and D. A. Lidar, Phys. Rev. Lett. {\bf 89}, 047901
 (2002).
\bibitem{lidar3} D. A. Lidar and L. A. Wu, e-print quant-ph/0302198

\bibitem{Ruskai}  M. B. Ruskai, Phys. Rev. Lett. {\bf 85}, 194 (2000).

\bibitem{Kane}  B. E. Kane, Nature {\bf 408}, 339 (2000).

\bibitem{Vrijen}  R. Vrijen, E. Yablonovitch, K. Wang, H. W. Jiang, \newline
A. Balandin, V. Roychowdhury, T. Mor, and D. DiVincenzo, Phys.
Rev. A {\bf 62}, 012306 (2000).

\bibitem{Zhou} X. X. Zhou, Z. W.Z hou, G. C. Guo, and M. J. Feldman, Phys. Rev. Lett. {\bf 89}, 197903 (2002)

\bibitem{Benjamin} S. C. Benjamin and S. Bose, Phys. Rev. Lett. {\bf 90}, 247901 (2003)

\bibitem{Zhang} Y. Zhang, Z. W. Zhou, B. Yu, and G. C. Guo, J. Opt. B: Quant Semi.Opt. {\bf 5}, 309 (2003).

\bibitem{DiVicenzo} D. P. DiVincenzo, Phys. Rev. A {\bf 51}, 1015 (1995)

\bibitem{Bremmer}  M. J. Bremner, C. M. Dawson, J. L. Dodd, A. Gilchrist, A. W. Harrow,
D. Mortimer, M. A. Nielsen, and T. J. Osborne, Phys. Rev.
Lett.{\bf 89}, 247902 (2002).

\end{references}
\end{document}